\documentclass[aps,prl,reprint,lengthcheck]{revtex4-2}

\usepackage{xcolor}

\usepackage{amsmath}
\usepackage{bm}
\usepackage{graphicx}
\usepackage{subcaption}
\usepackage{etoolbox}
\usepackage{tabularx}
\usepackage{multirow}
\usepackage{booktabs}
\usepackage{units}
\usepackage{gensymb}
\usepackage[font=small,labelfont=bf,
   justification=justified,
   format=plain,singlelinecheck=false,justification=raggedright]{caption} 
\usepackage{mathtools}

\usepackage{filecontents}

\begin{document}
\title{Frequency clustering and disaggregation in idealized fractal trees}
\author{F. Danzi}
\email{fdanzi@purdue.edu}
\author{J. M. Gibert}
\email{jgibert@purdue.edu}
\affiliation{Advanced Dynamics And MechanicS Lab,\\ School of Mechanical Engineering and Ray W. Herrick Laboratories, Purdue University, West Lafayette, (IN) USA  47907}

\begin{abstract}
The pattern of formation of resonant frequency clusters in idealized sympodial dichasium trees is revealed by numerical modeling and analysis. The largest cluster's cardinality correlates with that of a Small World Network which shares the same adjacency matrix. Compartmentalization of the modal characteristics and robustness to perturbations to the limb geometry are both dictated by topology and inherent symmetry of the structure and are not bound to specific allometry. When the spatial symmetry of the limb geometry is perturbed, we see the percolation of the largest cluster.
\end{abstract}

\maketitle

Fractals are ubiquitous in science as they adequately represent the geometry of many biological systems featuring self-similarity~\cite{Mandelbrot636}. Examples include micro-filament in the cytoplasm of eukaryotic cells~\cite{Sadegh2017}, cardiovascular \cite{Hahn2005} and bronchial system~\cite{Horsfield1971,Andrade1998}, magma conduit system \cite{CORTINI1990}, the wrinkled surface separating turbulent from non-turbulent regions in open shear flows ~\cite{deSilva2013}, proteins \cite{Enright2005}, clouds and rainfall areas~\cite{LOVEJOY185}, nucleotide sequence~\cite{Peng1992}, trees~\cite{Rodriguez2008} and basins river~\cite{Rodriguez1997,Montgomery1998}  and many other type of networks \cite{Song2005}. In modeling self-similarity (self-affinity), power-law such as allometry are used in constructing the theoretical foundation for the biometric study of biological systems~\cite{PhysRevE.72.036101,Snell1892,kendall_1977}. Self-similarity arises spontaneously in plants in response to hostile environmental conditions (thigmomorphogenesis~\cite{Moulia2006}). This phenomenon ultimately affects the power law which dictates the scaling of the limb geometry to enhance the response to wind~\cite{EloyPRL} and competition for light~\cite{Eloy2017} in trees.

Complete binary trees were used to study the influence of the branching angle in sympodial dichasium trees and mode localization enabling motionless trunk were theoretically and experimentally observed in sympodial dichasium trees with first and second-order branches~\cite{Kovacic_2018,Kovacic20181,Kovacic2020}, and in chains consisting of coupled pendulum or block masses attached mutually with springs~\cite{Kovacic2020}. Numerical analysis unveiled biological tuning to modal compartmentalization for sympodial trees with branches following Da Vinici's law~\cite{Rodriguez2008}. To date, observations on the effect of branching on the modal compartmentalization were obtained neglecting the stochastic nature of the scaling law and, assuming that limbs' geometry scales with Da Vinci's law.  

Based on an analytical derivation of the equations of motion, we propose a generalized model for studying the dynamics of open-growth fractal trees. The emphasis is on trees that have sympodial dichasium branching. The generalization enables us to understand the compartmentalization or clustering of the modal frequencies of trees into groups with nearly identical values. The effect of deviations from the allometry is considered by modifying the allometry's parameters or by introducing random perturbations to the limb geometry dimensions. In this way, we deconstruct the role of allometry as a means of biological tuning to modal compartmentalization in favor of tree (network) architecture symmetry and topology. In this paper, topology refers to the schematic arrangement of the tree's branches. 
Furthermore, bringing the discussion into the fold of symmetry and topology helps explain the spatio-temporal distribution of energy as well as the limb geometry's resilience (robustness) to damage.

\begin{figure}[htbp]
\centering
  \includegraphics[scale=.8]{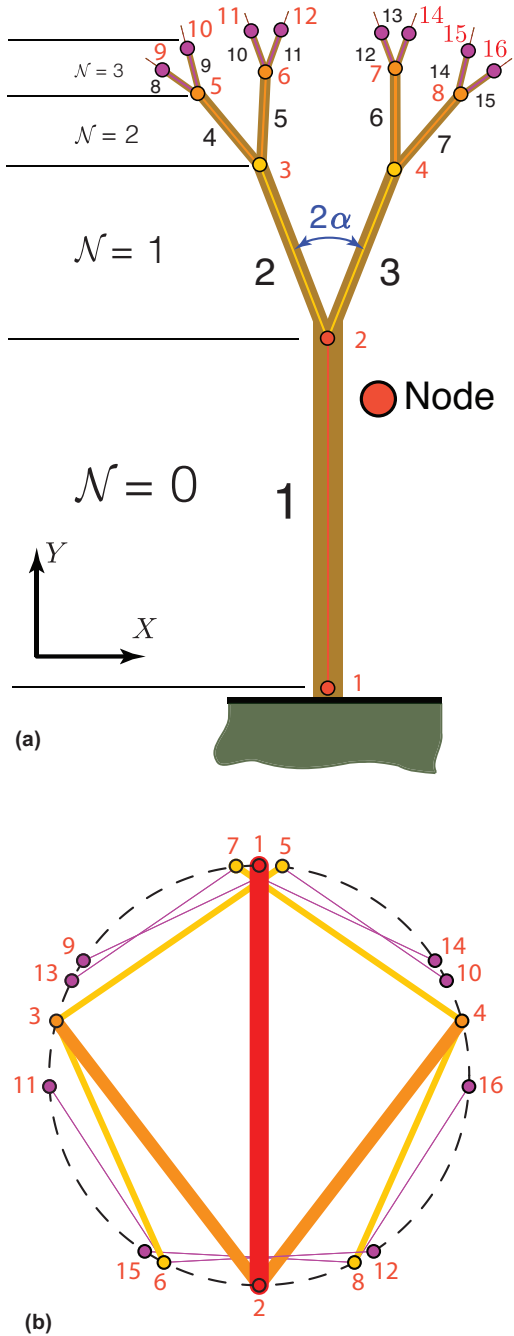}
\caption{(color online). Idealized tree architecture considered in this study with graph representation superimposed on the structure (a) and, (b) equivalent topology mapped on a ring. In (a) numerals in red indicate nodes, numerals in black indicate links or branches. }
\label{fig:SympodialTree}
\end{figure}

We consider a two-dimensional (2D) branched structure consisting of $2^{\mathcal{N}+1}-1$ branches where $\mathcal{N}$ is the branching level ($\mathcal{N}=0$ corresponds to the trunk). The sympodial dichasium tree shares the same architecture as a complete binary tree graph. In defining the tree as a graph, the branches are labeled in ascending order from left to right and from bottom to top as shown in Fig.\ref{fig:SympodialTree} (a). The allometry laws ~\cite{Rodriguez2008} can be written as
\begin{align}
D_{\mathcal{N}}/D_{\mathcal{N}-1}\simeq\sqrt{\lambda},~\text{and}~
L_{\mathcal{N}}/L_{\mathcal{N}-1}\simeq\lambda^{1/(2\beta)},
\end{align}
where $D$ represents the diameter of the branch,~ $L$ is the length of the branch, $\lambda$ is the lateral branching ratio and, $\beta$ is the branch slenderness allometric coefficient.
The topology is time invariant and the $i^{\textit{th}}$ node position is defined through the coordinate $(x_i,y_i)$ given by the following recursive law.

\begin{subequations}
\begin{align}
x_i=&x_j + L_t \lambda^{\mathcal{N}/(2\beta)} \cos(\varphi_i),\\
y_i=&y_j + L_t \lambda^{\mathcal{N}/(2\beta)} \sin(\varphi_i),
\end{align}
\end{subequations}
where $L_t$ is the length of the trunk and, $\varphi$ is the branching angle which can be expressed as
\begin{equation}
\varphi_i = \begin{cases}
\pi/2 \quad &\text{for} \quad  \mathcal{N}=0,\\
\varphi_j + (-1)^i \varphi \quad &\text{for} \quad  \mathcal{N}\ge1,
\end{cases}
\end{equation}
where $\varphi$ is the initial branching angle between the trunk and the first order branches. Defining $B=[2^{\mathcal{N}-1}+1,2^{\mathcal{N}}]$, $j$ is given as
\begin{equation}
j=
\begin{cases}
1 \quad &\text{for} \quad  \mathcal{N}=0,\\
j\in (B,2)\quad &\text{for} \quad \mathcal{N}\ge1. \end{cases}
\end{equation}
Here, $(B,2)$ is a 2-tuple where any element of the set $B$ has multiplicity of 2; finally, $i$ is defined as
\begin{equation}
i=
\begin{cases}
2 \quad &\text{for} \quad \mathcal{N}=0,\\
i\in [2^{\mathcal{N}}+1,2^{\mathcal{N}+1}]\quad&\text{for} \quad \mathcal{N}\ge1 .\end{cases}
\end{equation}
Each branch is modeled as a prismatic, isotropic, homogeneous Euler Bernoulli beam, i.e., the shear deformation is assumed negligible. The individual beam transverse displacement is represented using a third-order polynomial. The analysis is limited to infinitesimal deformations. In analyzing the tree's modal characteristics, the effect of (a) gravity, (b) foliage/crown, and, (c) aeroelastic interactions are neglected. Continuity of axial $u_r(\xi_r,t)$ and transverse $w_r(\xi_r,t)$ deformations in a given branch is enforced using the following recursive law~\cite{danzi2018generalized}
\begin{subequations}
\label{eq:kinetic_gen}
\begin{align}
w_r (\xi_r,t)&=\Biggl(w^*_r (\xi_r,t)+\xi_r\sum_{m \in \mathcal{M}} \frac{\mathrm{d}w_m}{\mathrm{d}\xi_m}\bigg|_{ L_m}\Biggr)\cos(\varphi_i)\nonumber\\&+w_{r-1}(L_{r-1},t),\label{eq:contw}\\
u_r (\xi_r,t)&=\Biggl(w^*_r (\xi_r,t)+\xi_r\sum_{m \in \mathcal{M}}\frac{\mathrm{d}w_m}{\mathrm{d}\xi_m}\bigg|_{ L_m}\Biggr)\sin(\varphi_i)\nonumber\\&+u_{r-1}(L_{r-1},t)\label{eq:contu},
\end{align}
\end{subequations}
where $w^*_r(\xi_r,t)=(\xi_r^2- \xi_r^3/(3L_r))q_r(t)$, is the transverse displacement in a local reference frame, $\xi_r$ is the scalar position along the $r$-th branch, $q_r(t)$ is the generalized coordinate, and $\mathcal{M}$ is the set of all the branches traversed in a depth-first-search sense going from the trunk to the branch $r$.
The Lagrangian of the system is $\mathcal{L}=T-U$ where $T$ and $U$ are respectively the kinetic and potential energy of the system.
The equilibrium generalized displacements for a system in free vibration where the effects of gravity are assumed negligible can be  determined from
\begin{equation}
    \frac{\partial U}{ \partial q_r}\bigg|_{\boldsymbol{q_e}=0}=0.
\end{equation}
The  kinetic energy can be written as 
\begin{equation}
T\approx\frac{1}{2}\displaystyle{\small \sum_{r=1}^n\sum_{s=1}^n} m_{rs}\dot{\hat{q}}_r \dot{\hat{q}}_s,
\end{equation}
where $m_{rs}=m_{sr}$ are the coefficients of  the mass matrix $\mathbf{M}$ and are defined as 
\begin{equation}
m_{rs}=\frac{\partial^2 T}{\partial \dot q_r \partial \dot q_s }\bigg|_{\boldsymbol{q_e}=0}.
\end{equation}
where $\boldsymbol{q_e}$ is the vector of equilibrium generalized coordinates. Similarly, the elements stiffness matrix $\mathbf{K}$ can be defined as
\begin{equation}
k_{rs}=k_{sr}=
\frac{\partial^2 U}{\partial q_r \partial q_s }\bigg|_{\boldsymbol{q_e}=0}.
\end{equation}
The architecture of the tree and the parameterization adopted leads to a hierarchical form of the stiffness and mass matrices which is detailed in the supplemental material~\cite{DanziPRLSupp} along with their sparsity pattern. In particular, we show that the elastic matrix has a repetitive $2 \times 2$ block matrix on the main diagonal, and a $2 \times 1$ vector off-diagonal that portrays the pairwise coupling among branches that are one branching level apart. The stiffness coefficients between two successive branching levels scale as $\tfrac{k^{\mathcal{N}+1}_{rs}}{k^{\mathcal{N}}_{rs}}\propto \lambda^{2+\tfrac{1}{2\beta}}$. The mass matrix has a different sparsity pattern w.r.t. the stiffness matrix. In particular, all the branches whose node(s) that are traversed moving from a given node to the root node are inertially coupled. The inertial coupling largely affect the dynamical response of the system. 
The linearized Lagrange equations yield a set of $N$ linear, constant-coefficient differential equations, the general form of the solution can be written as $\hat{\boldsymbol{q}}=\mathbf{\Phi}e^{\Lambda t}$  leads to the eigenvalue problem
$ (-\Lambda^2\mathbf{M}+\mathbf{K})\mathbf{\Phi}=\mathbf{0}$, where $\Lambda$ is an eigenvalue and the $\mathbf{\Phi}$ is the associated eigenvector. The number of degrees of freedom equals the number of branches, i.e. $n=2^{\mathcal{N}+1}-1$. We developed a numerical code based on the aforementioned analysis and the details provided in the supplemental material.

The dimensional and allometry parameters used in the following analysis are given as in~\cite{Rodriguez2008}: trunk of length $L_T=6.9$ m, branch slenderness allometric coefficient $\beta=3/2$, lateral branching ratio $\lambda=1/2$, branching angle $\varphi=20\degree$, Young modulus $E=11.3$ GPa, Poisson ratio $\nu=0.38$ and density $\rho=805$~kg/m$^3$. For simplicity, the Young modulus and the density are kept constant for all branching levels as in~\cite{Rodriguez2008}.
Without a loss of generality, the tree modeled here has $\mathcal{N}=3$ branching levels corresponding to 15 degrees of freedom (results pertaining to trees with higher-order branches and up to level branching $\mathcal{N}=6$ - i.e. 127dof - are reported in the supplemental material~\cite{DanziPRLSupp}).

\begin{figure*}[htbp]
\begin{center}
  \includegraphics[scale=1]{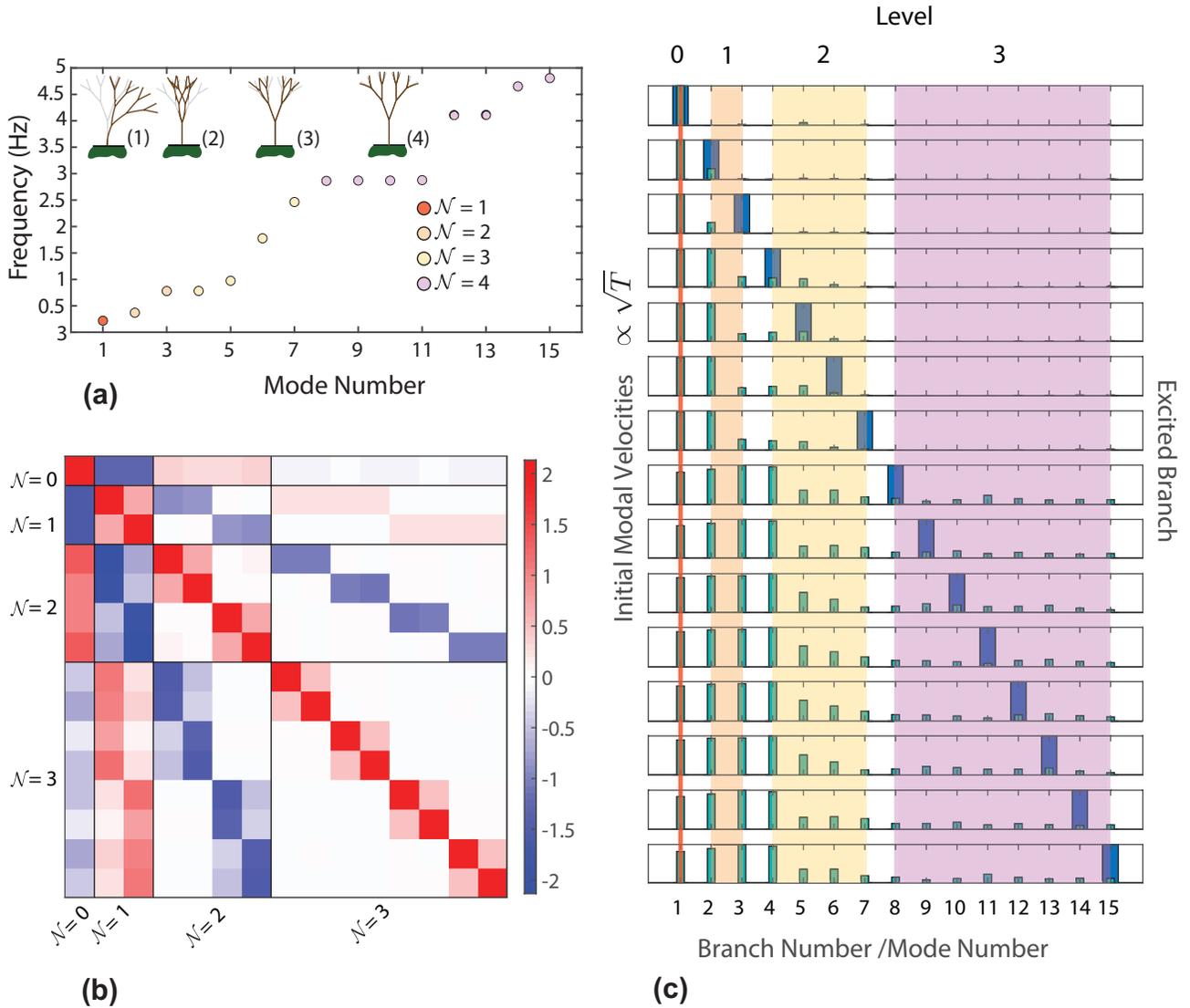}
\end{center}
\caption{(color online).
(a) Mode shapes and frequencies for a sympodial dichasium tree at branching level $\mathcal{N}=3$. The mode shapes of the unperturbed tree depicted here are associated to, respectively: (1) $f_1$ ($\mathcal{N}=0$), (2) $f_2$ ($\mathcal{N}=1$), (3) $f_4$ ($\mathcal{N}=2$) and, (4) $f_{10}$ ($\mathcal{N}=3$). The figure shows the clustering of the modal frequencies.
(b) Visualization of dynamical matrix ($\mathbf{B}=\mathbf{M}^{-1}\mathbf{K}$) of a sympodial dichasium tree at level branching $\mathcal{N}=3$. The matrix is nondimensionalized to have all the diagonal terms equal to 1. Notice the dominance of the upper diagonal matrix (i.e. coupling between upper branching levels with lower levels) w.r.t. to lower diagonal matrix.
(c) Initial modal velocity and excited branch geometry governed by da Vinci's allometry law. Bars in green denote the normalized initial velocity. Bars in blue denote the excited branch.  }
\label{fig:EnergyTree}
\end{figure*}

To understand the pattern of cluster formation, we compared the tree to a Small World Network (SWN) ~\cite{Watts1998}. To foster the analogy between the two representations, we mapped the tree on a ring with a diameter equal to the trunk length\footnote{Notice that this choice is deliberate and one could have assumed a circle with unit diameter without affecting the SWN representation of the idealized tree.}(see Figure~\ref{fig:EnergyTree} (b)). At a given level, the positions of the two child nodes are the two intersections on the circle with a euclidean distance equal to $\lambda^\frac{1}{2\beta}$ from the parent node. Contrary to the canonical representation of the SWN, the nodes calculated as above are unevenly spaced on the ring. Examining the results of the modal analysis we found that clusters with cardinality larger than one correspond to deformation of members at branches at $\mathcal{N}=2$ and continue for higher branching levels. Repeated eigenvalues in the cluster are due to eigenvalues with an algebraic multiplicity greater than one. In addition to the largest cluster, whose cardinality is $2^{\mathcal{N}-1}$ resonators, there are two additional clusters with cardinality equal to $2^{\mathcal{N}-2}$. This pattern is invariant with respect to the allometry law adopted (see supplemental material~\cite{DanziPRLSupp}) and persist for trees with branching level $\mathcal{N}\ge2$ . Furthermore, the following rule holds

\begin{equation}
C_{max} \approx (1-C) (2^{\mathcal{N}+1}-1),     
\end{equation}
where, $C$ is the clustering coefficient of the unweighted graph (or equivalently of the Small World Network depicted in Figure~\ref{fig:SympodialTree} (b))  and $C_{max}$ is the cardinality of the tree's biggest cluster.

The mode shapes associated with frequencies higher than the fundamental are localized in the higher branching levels. In particular, larger displacements are confined within a particular branching level while the other branches experience minimal bending deformation as depicted in Figure~\ref{fig:EnergyTree} (a). Modes' compartmentalization pattern reflects the clustering seen in modal frequencies. The mode shapes associated to the first ($\mathcal{N}=0$), third ($\mathcal{N}=1$), fifth ($\mathcal{N}=2$) and eighth ($\mathcal{N}=3$) frequencies respectively are reported in Figure~\ref{fig:EnergyTree}(a) (1-4). Other than numerically, we observed mode compartmentalization also experimentally as detailed in the supplemental material~\cite{DanziPRLSupp}.

Recognizing the relationship between topology and modal shapes, it is instructive to investigate the effect of topology on energy distribution in the tree. Figure~\ref{fig:EnergyTree} (c) shows a bar chart of the excited mode and initial modal velocities (proportional to square root modal energy in the absence of damping). The green bars correspond to the energy in a given mode and the blue bar corresponds to the branch being excited. Different branching levels are designated by different colored sections in the bar plot. Figure~\ref{fig:EnergyTree} (c) shows the initial energy when the tree branches scale with da Vinci's allometry law.  It is evident that the topology influences the initial energy distribution. Furthermore, if a given branch on level $\mathcal{N}$ is excited then the lower branch levels contain more energy than higher branch levels. These trends are evident regardless of the allometry law adopted to describe the limb geometry (see Figure~\ref{fig:EnergyTree} and Supplemental material~\cite{DanziPRLSupp}).  The allometry law does allow for the higher-level branches to have slightly more initial velocity but the overall distribution is similar. This agrees to what is anticipated by the Constructal Law for other types of branched systems~\cite{Bejan2000}. The localization of the modal deformation results in a structure in which energy flows predominantly unidirectionally regardless of the branch excited, resembling a mechanical diode. Mathematically, this behavior can be understood by examining the dynamical matrix of the system, $\mathbf{B}=\mathbf{M}^{-1}\mathbf{K}$, depicted in Figure~\ref{fig:EnergyTree} (b). Each row is the discretized equation of motion of one oscillator where the off-diagonal terms are the coupling between the i-th dof and the j-th dofs, with $j\neq i$. A cursory inspection of the dynamical matrix reveals two aspects: 1) the matrix is non-hermitian, and 2) the lower diagonal terms are, at least in magnitude, larger than the lower diagonal terms, i.e., $\left|d^{j\le i}_{i,j}/d^{j> i}_{i,j}\right|>>1$.  This implies that the dynamic of the i-th dof is predominantly affected by all the branches belonging to lower branching levels. Results pertinent tree with higher branching levels show the same trend and are reported in the supplemental material. Another key aspect that can be revealed looking at the dynamical matrix is that $\bf{B}$ is a block centrosymmetric matrix regardless of the branching levels that form the tree; this leads to the clustering of the modal characteristics~\cite{ANDREW1973151}.

\begin{figure*}[htbp]
\centering
  \includegraphics[scale=1]{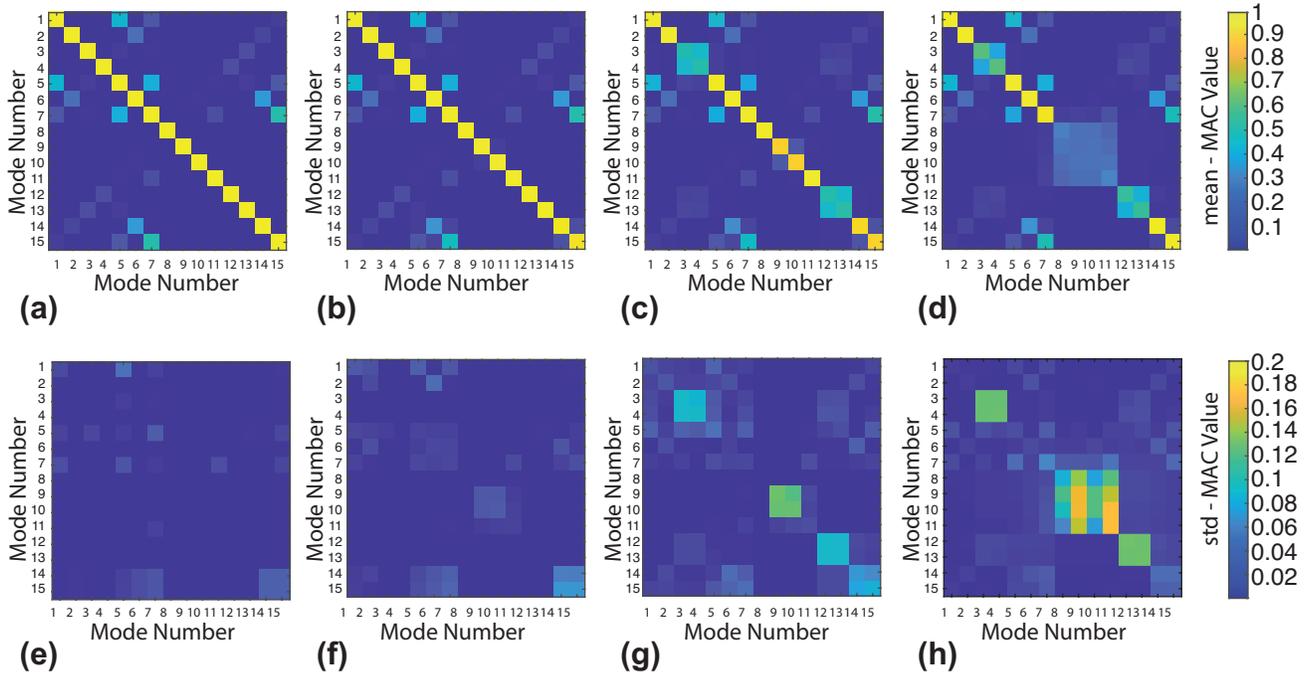}
\caption{Study of the resilience of the cluster for stochastic trees with Gaussian random perturbations applied to the branching levels in red on the inset image. Markers in (a)-(d) denote the frequencies of the tree with unperturbed allometry laws, the error bars show the variability of the frequencies on 1000 realizations. Perturbations are applied to: (a) the trunk, (b) $\mathcal{N}=1$, (c) $\mathcal{N}=2$ and, (d) $\mathcal{N}=3$. Mean Value of the Modal Assurance Criterion (MAC) for the perturbed tree on 1000 realizations; perturbations to: (e) the trunk, (f) $\mathcal{N}=1$, (g) $\mathcal{N}=2$ and, (h) $\mathcal{N}=3$. Standard deviations of the MAC when perturbations are applied to: (i) the trunk, (j) $\mathcal{N}=1$, (k) $\mathcal{N}=2$ and, (l) $\mathcal{N}=3$. Notice how, when perturbations are applied to the branching level $\mathcal{N}=3$, the standard deviation spikes in the $4\times4$ minor that corresponds to the largest cluster, therefore, eliciting percolation.}\label{fig:MAC}
\end{figure*}

\begin{figure}
    \centering
    \includegraphics[scale=1]{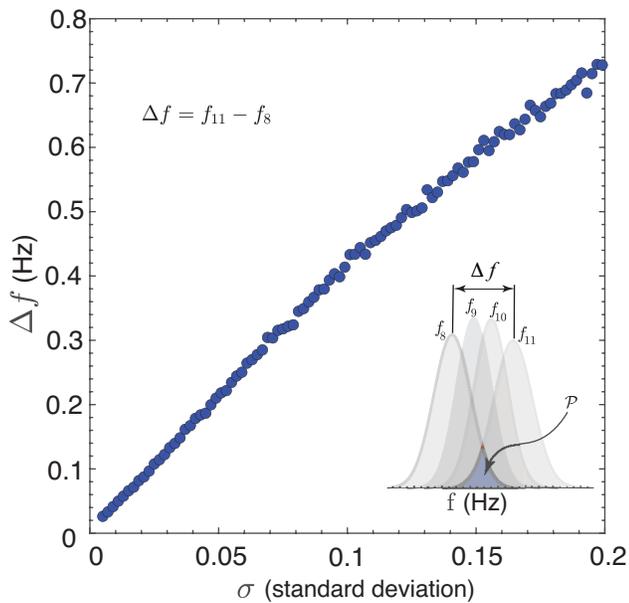}
    \caption{Set-space $\Delta f_c$ (Hz) of the frequencies of the largest cluster as a function of the perturbation $\sigma$ applied to branches of the top level ($\mathcal{N}=3$). The figure shows that as the perturbation increases the set-space of the frequencies varies $\Delta f_c\le0.01 Hz$ to $\Delta f_c\approx1Hz$. In the limit, the probability of existence of the largest cluster, i.e. the area enclosed by the statistical distributions of $f_8$ and $f_{11}$, is $\mathcal{P}=0.13.$}
    \label{fig:Delta-sigma}
\end{figure}

The robustness of the cluster, i.e., the effect of changes in the limb geometry on the group of resonance frequencies and mode shapes, is exemplified in Fig.~\ref{fig:MAC} (a) to (h) where deviations of the branches' diameters from the allometry law are related to the variations of the correlation between the mode shape of the perturbed and unperturbed tree. Although field measurements show that multivariate statistical distribution is required to describe limb dimensions in real trees~\cite{Hafley1977}, here, without loss of generality,  we considered that the branches' diameter follows a normal distribution. The analysis here is limited to variations of branches diameter, results concerning variations of the lengths and angle are reported in the supplemental material. We assigned randomly generated scaling factors $r$ (mean $\bar{\mu}=1$, standard deviation $\bar{\sigma}=0.05$) to all the branches at one level and repeated the numerical experiment on 1000 realizations. Denoting with $D_{\mathcal{N}}$ the branches' diameter , the perturbed diameter $D^*_{\mathcal{N}}$ is, $D^*_{\mathcal{N}}=r(\mu,\sigma) D^*_{\mathcal{N}}$, where $r$ is the scaling factor. Due to the perturbations in the limb geometry, the structure is no longer symmetric implying a fragmentation of the clusters.
To have a qualitative measure of the clusters' robustness, we project the perturbed eigenvectors onto the unperturbed vector space. Albeit, the frequencies of the perturbed and unperturbed systems are not coincident, their variation is limited if the perturbation is small; this justifies the use of the Modal Assurance Criterion (MAC~\cite{Allemang1982ACC}) as a correlation metric. 
The MAC is defined as:
\begin{equation}
\text{MAC}= \frac{\left|\mathbf{\Phi}^T_i\mathbf{\Phi}_i\right|^2}{\mathbf{\Phi}^T_i\mathbf{\Phi}_i\mathbf{\Phi}^T_j\mathbf{\Phi}_j}
\end{equation}
Here, $\mathbf{\Phi}_i$ and $\mathbf{\Phi}_j$ are the eigenvectors of the perturbed and unperturbed system respectively; the superscript $T$ denotes transpose. The mean value of the MAC associated with 1000 realizations is depicted in Figure~\ref{fig:MAC} (a)-(d). The projection of the perturbed vector space into the unperturbed one is consistent with the projection of the unperturbed vector space into itself. The system preserves its diagonal dominance; therefore, the qualitative response of the perturbed tree resembles that of the unperturbed one. Moreover, the standard deviations (Figure \ref{fig:MAC} (e)-(h)) on 1000 realizations is limited meaning that the eigenvectors' space is marginally affected by the perturbations.  This result  emphasizes the resilience of the cluster. It is worth noticing that, when perturbations are applied to the branching level $\mathcal{N}=3$, the standard deviation spikes in the $4\times4$ minor enclosed by row/column 8 to 11, this region corresponds to the location of the largest cluster, therefore, triggering percolation of the cluster. To further understand this behavior and to investigate the effect of deviations from the allometry law, we let the standard deviation vary and track the set-space of the frequencies of the largest cluster $\Delta f_c = f_{11}-f_8$ on 1000 realizations. We noticed an almost linear trend between the standard deviation of the scaling factor of the branches $\sigma$ and the frequencies set-space $\Delta f_c$ (Figure~\ref{fig:Delta-sigma}). When the perturbation increases, the set-space becomes large enough that the frequencies are distinguishable therefore eliciting the dis-aggregation of the cluster. Calculating the probability of the existence of the largest cluster as the overlapping area between the statistical distributions of $f_8$ and $f_{11}$, we saw that if the perturbation exceeds a threshold value, the probability settles at $0.13$; for values of the standard deviation under the threshold value, the frequencies' set-space is so small that the four frequencies are almost coincident and the probability of having the cluster with the maximum cardinality tends to 1. We also noticed that, when the perturbations to the limb geometry increase, there is a non-negligible probability that the largest cluster would disappear in favor of smaller clusters. In some cases, frequencies that were not clustered will group as a result of large perturbations applied to the limb geometry, therefore, implying a redistribution of the clusters. Qualitatively, the same conclusion can be drawn perturbing branches lengths. In contrast, we noticed that small perturbations applied to the branching angle do not affect the cardinality of clusters nor the correlation of the perturbed and uperturbed mode shapes (see~\cite{DanziPRLSupp}).

This letter, in summary, introduces a hierarchical model of an idealized sympodial dichasium fractal tree. The model was used to reveal the formation pattern of modal frequency clusters as well as their cardinality. The results presented here offer a simple yet compelling picture: clustering  and localization of the modal characteristics, and the robustness of trees to perturbation of the limb geometry are universal properties related to the topology of the network and not bound to particular allometry. The cardinality of the largest cluster can be determined by calculating the system's clustering coefficient of the associated Small World Network. Clusters are resilient to perturbations up to a threshold value before percolating. Energy flows nearly unidirectionally in the system resembling a mechanical diode. 

The authors would like to acknowledge the financial support provided by the NSF CAREER Award: CMMI 2145803 and the Purdue Research Foundation (PRF). We thank Hongcheng Tao (Advanced Dynamics And Mechanics Lab, Purdue University) and Dr. Myungwon Hwang (Adaptive Structures Lab, Purdue University) for the assistance provided for the experiments.

\bibliographystyle{apsrev4-1}
\bibliography{Ref_PRL_Danzi_Tree} 

\end{document}